\documentclass[twoside]{article}
\usepackage{fleqn,espcrc2,epsfig}

\usepackage{graphicx}

\usepackage[figuresright]{rotating}

\newcommand{\be}{\begin{equation}}
\newcommand{\ee}{\end{equation}}
\newcommand{\bea}{\begin{eqnarray}}
\newcommand{\eea}{\end{eqnarray}}

\newcommand{\AmS}{{\protect\the\textfont2
  A\kern-.1667em\lower.5ex\hbox{M}\kern-.125emS}}

\title{The interaction between center monopoles in $SU(2)$ Yang--Mills}

\author{Philippe de Forcrand\address[ETH]{Inst. f\"ur Theoretische
    Physik, ETH H\"onggerberg, CH-8093 Z\"urich, Switzerland}
\address{CERN, Theory Division, CH-1211 Gen\`eve 23, Switzerland}, 
Massimo D'Elia\address{Dipartimento di Fisica dell'Universit\`a
  and I.N.F.N., I-56127, Pisa, Italy} and Michele
Pepe\addressmark[ETH]\thanks{Presented by M. Pepe at LAT00.}}

\begin{document}

\begin{abstract}
We study the potential between a static center monopole and antimonopole
in $4d$ $SU(2)$ Yang--Mills theory. Using a new numerical
method, we show that the 't~Hooft loop is a dual order parameter
with respect to the Wilson loop, for the deconfinement phase transition.
We observe a $3d$
Ising--like critical behaviour for the dual string tension related to
the spatial 't~Hooft loop as a function of the temperature.

\end{abstract}

\maketitle

\section{Introduction}

In the $4d$ $SU(2)$ Yang--Mills theory a second order phase transition 
occurs. A critical temperature $T_c$ separates a cold confined phase
and a hot deconfined one. More than two decades ago, 't~Hooft 
proposed\cite{tHooft1} that the center degrees of freedom 
might play an important role in the confinement
mechanism. Along with this approach, he introduced an operator --
named 't~Hooft loop -- with a non trivial structure with respect to
the center subgroup, and argued that it could be used as a dual order parameter 
for the deconfinement phase transition.
The 't~Hooft loop, $\tilde{W}(C)$, is associated with a
given closed contour $C$, and is defined in the continuum $SU(N)$ theory by the 
following equal--time commutation relations\cite{tHooft1}
\bea
\left[ W(C), W(C') \right]~=~\left[ \tilde{W}(C), \tilde{W}(C') \right]~=~0~ \\
\tilde{W}^\dagger(C) W(C') \tilde{W}(C)~=~e^{i \frac{2\pi}{N} n_{CC'}} W(C')
\eea
where 
$W(C')$ is the Wilson loop associated with the closed contour $C'$ and
$n_{CC'}$ is the linking number of $C$ and $C'$. 
Just like the Wilson loop creates an elementary electric flux along $C'$,
the 't~Hooft loop creates an elementary magnetic flux along the path $C$ 
affecting any Wilson loop ``pierced'' by $C$. 
In that sense, the two types of loop are dual to each other. 
At zero temperature, it has been shown\cite{tHooft1,Tomb,Sam} 
that also the 't~Hooft loop behaviour is dual to that of the Wilson loop:
in the absence of massless excitations, an area law for one implies 
a perimeter law for the other, and vice versa. 
Hence, at $T=0$ the 't~Hooft loop obeys a perimeter law.

Several analytical\cite{Kaltes2,Kaltes1} and numerical\cite{KandT,Rebbi,Delde,we}
studies have been carried out in order  
to investigate this issue of duality at finite temperature. At
$T>0$, the Lorentz symmetry is broken, so spatial and 
temporal loops can have different behaviours. Because the 
spatial string tension persists also above $T_c$ for the Wilson loop, 
temporal 't~Hooft loops are expected to show a perimeter law in both phases; 
spatial 't~Hooft loops, on the other hand, are expected to obey a perimeter law in the confined 
phase and an area law -- defining a dual string tension (strictly
speaking it is an action density) -- in the deconfined phase. 

Using a novel computational approach, we have performed\cite{we} a numerical
study of the 't~Hooft loop, showing that it has a dual behaviour
with respect to the Wilson loop both at $T=0$ and $T>0$. 
Our main result is that the 't~Hooft loop is indeed a dual order
parameter for the deconfinement phase transition. 

\section{The lattice formulation}

The 't~Hooft loop characterizes the static potential between a center
monopole and antimonopole, just like the Wilson loop does for two
electric charges. Thus, to obtain a 't~Hooft loop, one creates a static 
monopole -- antimonopole pair, then the two charges
propagate and, at the end, annihilate. This is the standard
definition\cite{Guth,Suss} of the 't~Hooft loop on the lattice.
Specifically, consider the $SU(2)$ lattice gauge theory with the usual
Wilson plaquette action,
\be
S(\beta) = \beta \sum_P (1 - \frac{1}{2} \mbox{Tr} (U_P))
\label{wilsact}
\ee
where the sum extends over all the plaquettes $P$ and $U_P$ is the
path--ordered product of the links around $P$. Starting from
$S(\beta)$, one defines the partition function 
\be
Z(\beta) = \int [dU] \exp(- S(\beta))
\label{wilspfc}
\ee
To insert a monopole -- antimonopole pair, one must create
a flux tube with non trivial value with respect to the
center degrees of freedom: the center monopoles lie at the two ends of the tube.
The flux is switched on by multiplying by a non trivial
center element $z$ the plaquettes along a path, in the dual lattice,
joining the two monopoles. For $SU(2)$, the only non trivial center
element is $-1$ and so multiplying a plaquette by $z$ is equivalent to flipping its
coupling. Replicating this construction at successive time--slices, we
create an elementary magnetic flux along a closed contour $C$ in the
dual lattice, extending in one space- and in the time-direction (temporal 't~Hooft
loop). In a similar way, we can make the closed contour $C$ extend
only in spatial directions (spatial 't~Hooft loop). The set $\cal
P(\cal S)$ of the plaquettes whose 
coupling is flipped, $\beta \to - \beta$, is dual to a surface $\cal S$ 
supported by $C$. The action $S_{\cal S}$ of the system where an
elementary flux along a closed contour $C$ has been switched on is
then given, up to an additive constant, by
\be
S_{\cal S}(\beta) = - \frac{\beta}{2} \left( 
\sum_{P \notin {\cal P(\cal S)}}
\hskip -2mm
\mbox{Tr} (U_P) - 
\hskip -3mm
\sum_{P \in {\cal P(\cal S)}}
\hskip -2mm
\mbox{Tr} (U_P)  \right) 
\label{modiact}
\ee
and the partition function is
\be
Z_C(\beta) = \int [dU] \exp(- S_{\cal S}(\beta))
\label{modipfc}
\ee
$Z_C(\beta)$ does not depend on the particular chosen
surface $\cal S$, since different choices are related by a 
change of integration variables. 

The expectation value of the 't~Hooft loop gives the free energy cost
to create the flux loop. It is obtained by comparison between the states 
with the loop switched on and the states where it is~off:
\be
\langle \tilde{W}(C) \rangle = 
Z_C(\beta) / Z(\beta) 
\label{tllatdef}
\ee
This expression can be rewritten in the form
\be
\langle \tilde{W}(C) \rangle = \langle \exp 
\big( - \beta \sum_{P \in {\cal P(\cal S)}}\mbox{Tr} (U_P) \big)\rangle 
\label{tlprac}
\ee
with the average taken with the standard Wilson action.
The numerical computation of the ratio (\ref{tllatdef}),
(\ref{tlprac}) is a very hard numerical task due to the very poor
overlap between the relevant phase space of the numerator and the
denominator. 

Recently, the 't~Hooft loop, or special cases of it, have been studied
numerically on the lattice. In \cite{KandT,Rebbi} the sampling problem 
was solved by using a multihistogram method. In our study, we adopt
a new approach, where the ratio  $Z_C(\beta) / Z(\beta)$
is rewritten as a product of intermediate ratios, each easily measurable. 

\section{The numerical method}

In this section we present the new numerical technique that has been
developed to measure the expectation value of the 't~Hooft loop
operator. The direct evaluation of $\langle \tilde{W}(C) \rangle$ with 
(\ref{tllatdef}), (\ref{tlprac})  by a single Monte Carlo simulation is not 
reliable. Importance sampling of $Z(\beta)$ leads to generating field
configurations whose contribution to $Z_C(\beta)$ is negligible
and vice versa. The usual approach to overcome this difficulty is the 
multihistogram method\cite{KandT,Rebbi}, where one performs several
different simulations in which the coupling of the stack of plaquettes
in $\cal P(S)$ is gradually changed from $\beta$ to $-\beta$.
Instead, we interpolate in the number of plaquettes in 
$\cal P(S)$ with flipped coupling. We make use of the following
identity, where $N$ is the total number of plaquettes belonging to 
$\cal P(S)$
\be
\frac{Z_C(\beta)}{ Z(\beta)} =
\frac{Z_N(\beta)}{ Z_{N-1}(\beta)} \cdot
\frac{Z_{N-1}(\beta)}{ Z_{N-2}(\beta)} \cdot\ldots
\cdot \frac{Z_{1}(\beta)}{ Z_0(\beta)}
\label{factor}
\ee
where $Z_k(\beta)$, $k=0,\dots , N$ ($Z_N\equiv Z_C$ and $Z_0 \equiv
Z$) is the partition function of the system
where only the first $k$ plaquettes in $\cal P(S)$ have flipped
coupling. Consider the following figure where $\cal P(S)$ is the set
of bold and thin plaquettes, whose coupling has to be flipped to
create the loop. We interpolate between $0$ and $N$ flipped plaquettes by
considering a snake--like movement where the coupling is progressively flipped. 
\vskip -.6cm
\begin{figure}[h]
\begin{center}
\epsfig{figure=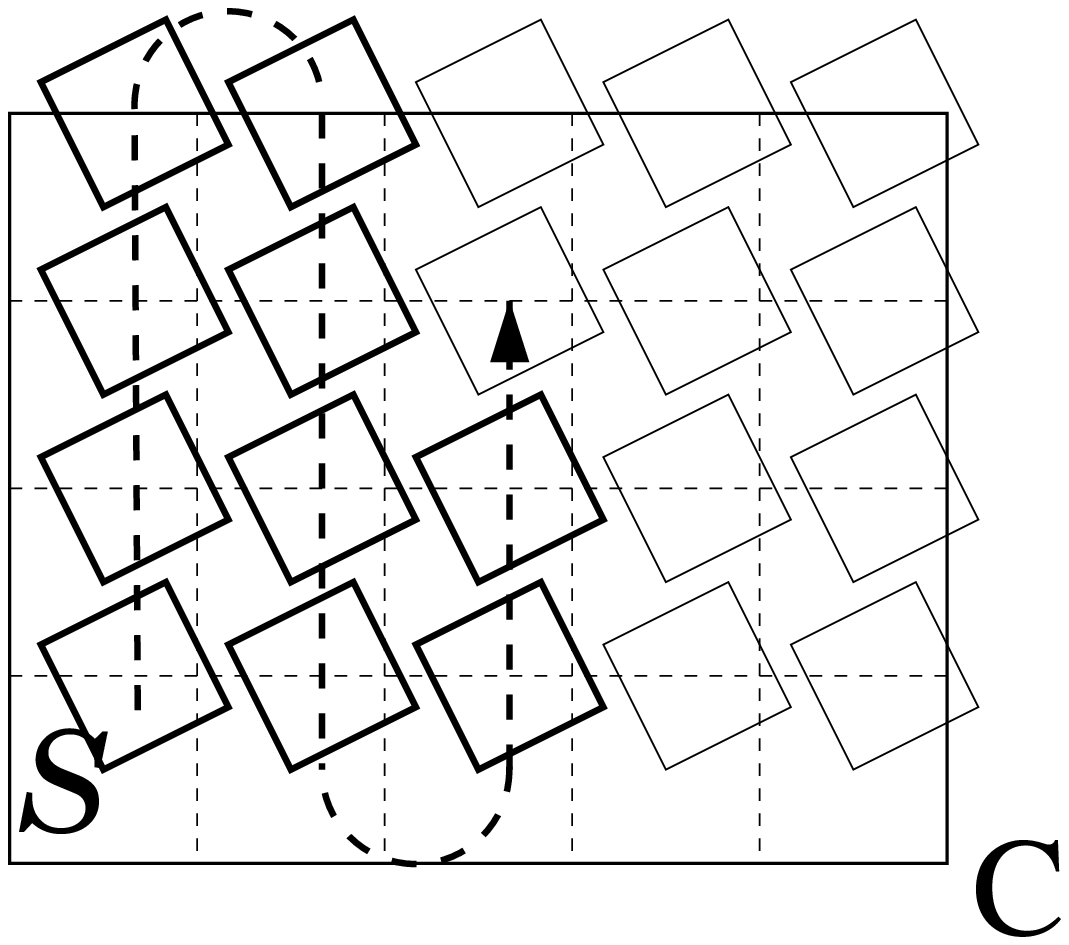,height=4.3cm,width=4.9cm}
\end{center}
\label{snake}
\end{figure}
\vskip -.9cm
Figure 1 shows the snake--like flipping movement: bold plaquettes have
coupling $-\beta$ and thin ones $+\beta$.
Thus, performing $N$ independent Monte Carlo simulations -- each 
corresponding to a different number of plaquettes in $\cal P(S)$ with
flipped coupling -- one can have a reliable estimation of 
$\langle \tilde{W}(C) \rangle$. The efficiency of the method
can be further enhanced through variance reduction tricks, described in detail
in \cite{we}.
An advantage of our method over the multihistogram technique 
is that the products in (\ref{factor}) give information on
smaller 't~Hooft loops for free; moreover, the error analysis is simpler
and less delicate than in a multihistogram analysis.

\section{Results}

We focus on the free energy $F(R)$ of a pair of static 
center monopoles as a function of their separation $R$. It can be obtained as 
${\rm lim}_{R_t \rightarrow \infty} -{\rm Ln}[\langle \tilde{W}(R,R_t) \rangle] / R_t$
by taking elongated $R \times R_t$ rectangular loops, in the same way as one
extracts the static potential between two chromo-electric charges. We take $R_t$
as large as possible, i.e. equal to the lattice size $L$.
This is analogous to measuring the correlation of two Polyakov loops, and is
the correct approach at finite temperature. Therefore we must flip the
coupling of $R \times L$ plaquettes. We do this with the snake--like
movement: we scan first the $R_t-$ and then the $R-$direction. 
With this ordering, the intermediate partition functions
$Z_L, Z_{2L},..,Z_{R \times L}$ in (\ref{factor})
provide us with the free energy at separations $1,2 .., R$ respectively.
The final ratio $Z_{L \times L}/Z_0$ gives the free energy of a center vortex
as computed in \cite{KandT}.

At zero temperature the 't~Hooft loop is expected to obey a perimeter law:
$Z_k / Z_0 \propto e^{-c \tilde{P}_k}$, where $\tilde{P}_k$ is the
length of the contour $C_k$. For the sequence of `snake' flipped
plaquettes -- up to boundary effects occurring for very
small or very large contours -- $\tilde{P}_k$ assumes only two different
values\cite{we}. Indeed numerical simulations show that $Z_k / Z_0$
centers around two values only, confirming the perimeter law. 
Furthermore, a direct measurement of Creutz ratios $\chi(R,R)$ shows a quick drop
to zero with the distance $R$.  

The free energy $F(R)$ can be fitted by a Yukawa form $\frac{e^{-m R}}{R}$,
up to an irrelevant additive constant. However the screening mass $m$ is
rather large, so that the signal quickly dies out. The measured value
of ca.~$2$ Gev for $m$ is close to the lightest gluonic excitation, the
scalar glueball ($\sim 1.65$ GeV), as observed in \cite{Rebbi,Delde}.
More precise statements would require a much more ambitious numerical study.

At finite temperature, the spatial and the temporal loops can have
different behaviours. This difference can be understood by the
following argument. Consider a spatial 't~Hooft loop, that is a loop
in the $(x,y)$ plane obtained by flipping the coupling of plaquettes in
the $(z,t)$ plane. The magnetic flux could diminish its free energy by spreading out,
but this is limited by the finiteness of the
$t$--direction. Conversely a temporal loop can spread its flux without
limitation. Thus we expect higher free energies for
spatial loops over temporal ones. Moreover, a dual behaviour of the
't~Hooft loop with respect to the Wilson loop leads to the following
scenario. The spatial Wilson loop obeys an area law below and above $T_c$
and so we expect a perimeter law behaviour, with a Yukawa term, for
the temporal 't~Hooft loop. The temporal Wilson loop obeys a perimeter
law above $T_c$ and so the spatial 't~Hooft loop should obey a
perimeter law, with a Yukawa term, below $T_c$ and an area
law above $T_c$, defining a dual string tension. 
This is precisely what we observe. 

\pagebreak
\begin{figure}[th]
\begin{center}
\epsfig{figure=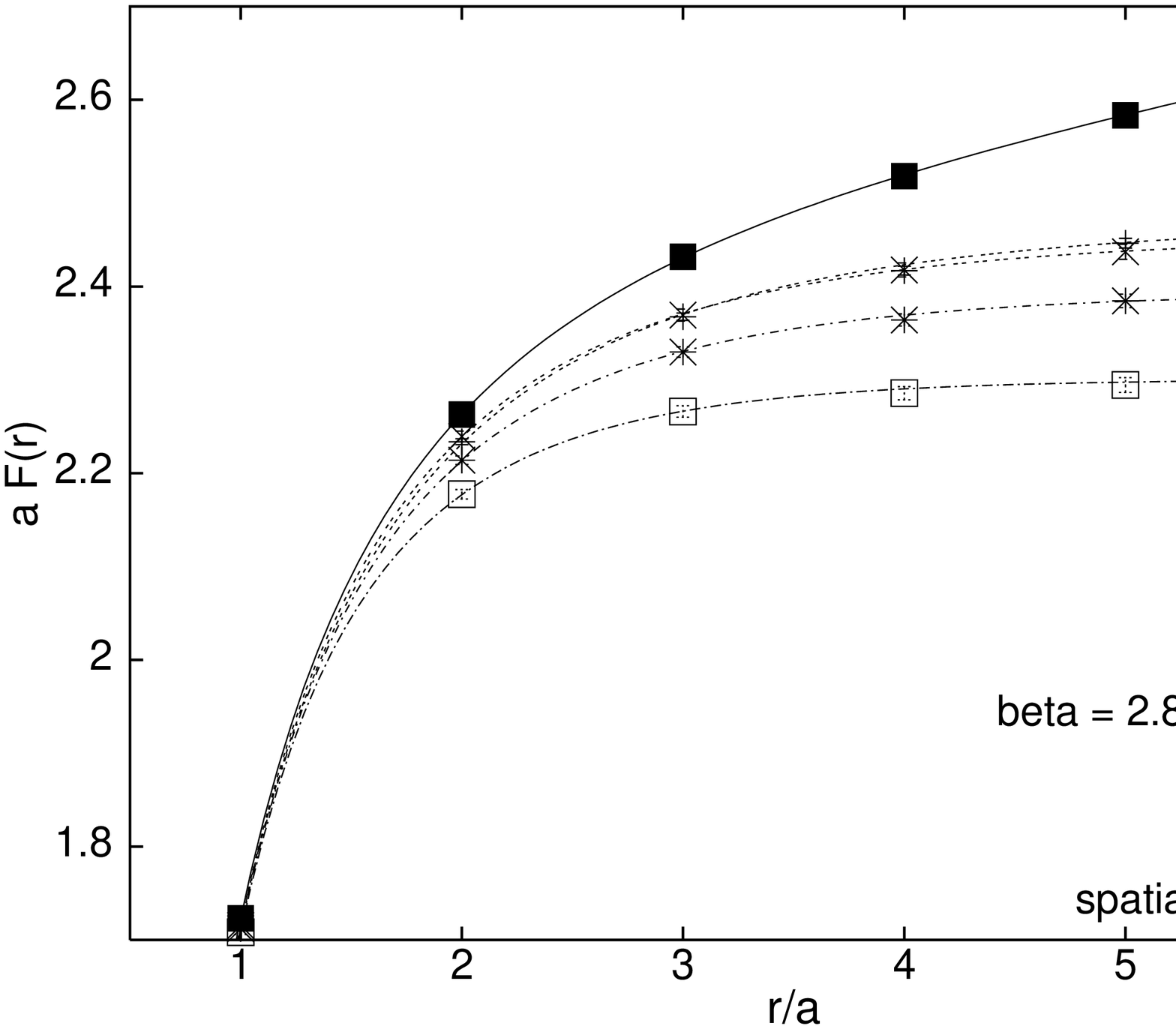,height=4.7cm,width=6.5cm}
\end{center}
\label{potential}
\end{figure}
\vskip -.9cm
Figure 2 displays our results for the free energy
vs. monopole separation at $T > T_c$. The spatial 't~Hooft loop (full
squares) shows a dual string tension; the temporal 't~Hooft loop (other data) shows a screening mass increasing with $T$. An attempt to fit the data for the temporal 
loop with the ansatz $F_0 + c \frac{e^{-m R}}{R} + \sigma R$, 
including a linear term, gives a dual string tension $\sigma$ consistent with zero. 
In contrast, this linear term is required to obtain an acceptable fit
above $T_c$ for the spatial loop: a dual string tension appears.
We have measured the screening mass $m$ at various temperatures both for
the spatial and the temporal 't~Hooft loops. In the former, it
seems to be little affected by temperature, while in the 
latter we observe a linear dependence in $T$, much like for the glueball
excitation which it presumably represents\cite{Datta}. As for $T=0$,
an accurate numerical study is in order to make precise quantitative  
statements. 

\vskip -0.5cm 
\begin{figure}[h]
\begin{center}
\epsfig{figure=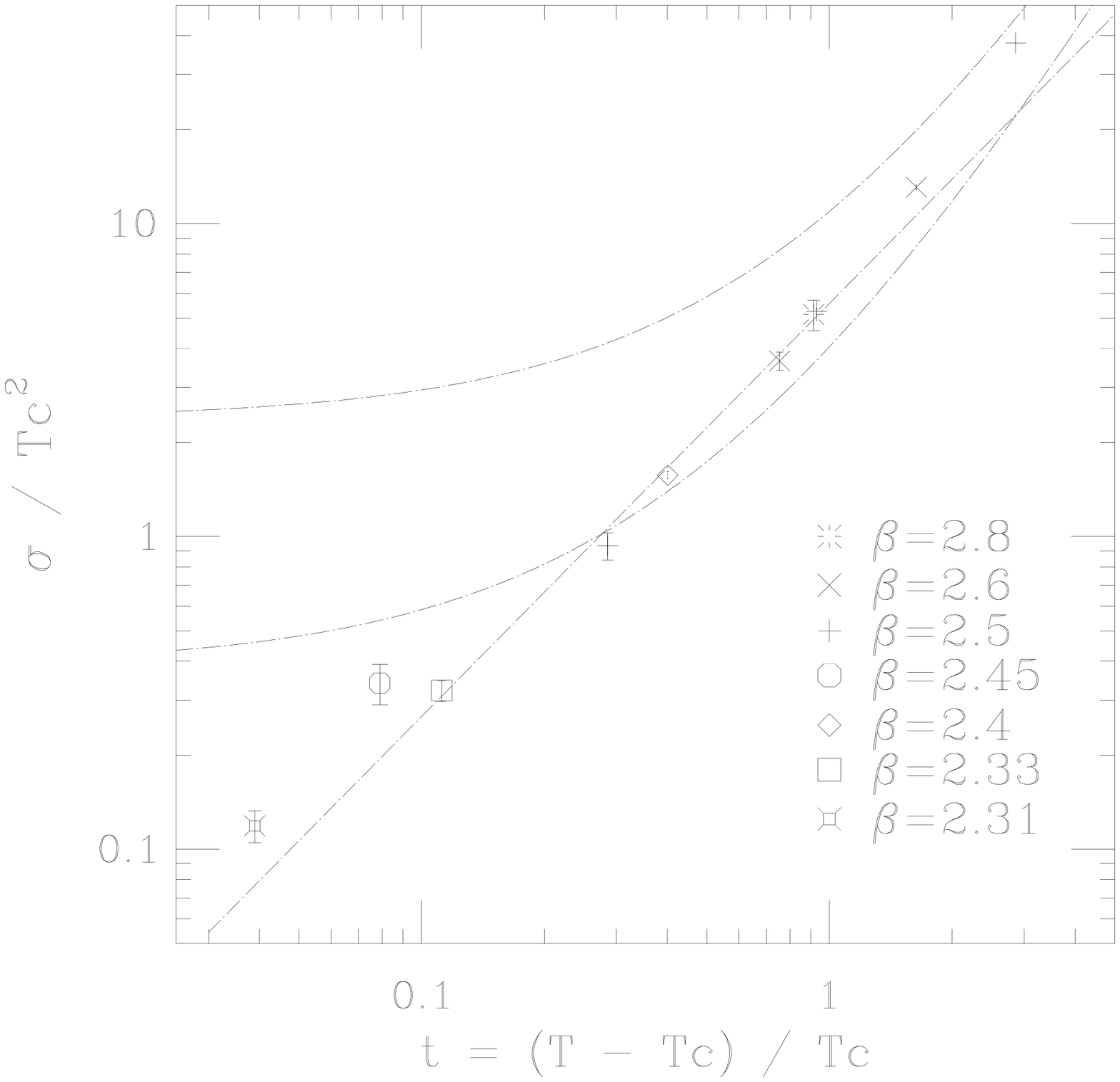,height=4.6cm,width=6.3cm}
\end{center}
\label{sigmavsT}
\end{figure}
\vskip -.9cm
The dual string tension $\sigma$ depends on temperature and must vanish
at $T_c$. 
Figure 3 shows that it does so as 
$\sigma \propto t^{2\nu}$, where $t = \frac{T-T_c}{T_c}$ 
is the reduced temperature.
The straight line is a power law fit to the $t < 1$~data, and the curves are
the perturbative result, to leading (upper) and next (lower) order.
The fitted critical exponent $\nu$, associated with the
correlation length $\xi = \sigma^{-1/2}$, comes out very close to that
of the $3d$ Ising model: $0.66(3)$ vs. $\approx 0.63$.
This should be expected since both models are in the same 
universality class. This dual string tension can then be taken as
order parameter for the restoration of the (magnetic) $Z_N$ symmetry,
corresponding to {\em de}confinement \cite{Kaltes2,Kovner}.

\section{Conclusions}

Using a new numerical method, we have shown that the 't~Hooft
loop is a dual order parameter for the deconfinement phase transition 
in $SU(2)$. We have observed a dual string tension 
in the deconfined phase with $3d$ Ising--like critical exponent.
We have measured the screening masses and studied their dependence on
the temperature; more accurate investigations are required for more
quantitative results on this point.

\end{document}